\documentclass[11pt]{article}
\usepackage[margin=1.25in]{geometry}
\pdfoutput=1
\usepackage[utf8]{inputenc}
\usepackage[english]{babel}
\usepackage[T1]{fontenc}
\usepackage{amsmath}
\usepackage{amssymb}
\usepackage{bbm}
\usepackage{physics}
\usepackage{xcolor}
\usepackage{mathtools}
\usepackage{hyperref}
\usepackage{parskip}
\usepackage{stmaryrd}
\usepackage{caption}
\usepackage{amsthm}
\usepackage{multirow}
\usepackage{float}

\usepackage{algorithm}
\usepackage{algpseudocode}
\usepackage{docmute}
\usepackage{caption}
\usepackage{subcaption}
\usepackage[symbol]{footmisc}
\theoremstyle{remark}

\theoremstyle{definition}
\newtheorem{defi}{Definition}

\newcommand\tildeB{{\stackrel{\sim}{\kern0pt B}}}

\usepackage{authblk}

\usepackage{tikz}
\usetikzlibrary{quantikz2}
\usetikzlibrary{positioning}
\usetikzlibrary{decorations.pathreplacing,angles,quotes}
\title{Lossless Reductions and Hybrid Sampling for Phylogeny Using the Maximum Compatibility Criterion}
\author[1,2]{Orson Ye\thanks{\texttt{ohy21@cam.ac.uk}}}
\author[3]{James Bonfield}
\author[1,2]{Joshua Cudby}
\author[4]{Richard Durbin}
\author[2]{Sergii Strelchuk}
\affil[1]{Department of Applied Mathematics and Theoretical Physics, University of Cambridge, Cambridge, United Kingdom}
\affil[2]{Department of Computer Science, University of Oxford, Oxford, United Kingdom}
\affil[3]{Wellcome Sanger Institute, Hinxton, United Kingdom}
\affil[4]{Department of Genetics, University of Cambridge, Cambridge, United Kingdom}
\date{}

\begin{document}

\maketitle

\begin{abstract}

Maximum compatibility provides a criterion for phylogenetic reconstruction in which the objective is to retain a maximum set of mutually compatible characters. For binary characters, pairwise compatibility can be represented by a weighted graph whose vertices correspond to characters and whose edges connect compatible pairs. A leading algorithm by Cherry [BMC Bioinformatics, 18(1):127, 2017] uses this representation to identify maximum compatible character sets as maximum weight cliques. 

We formulate the problem as maximum weight independent set enumeration on the complement graph and use deterministic postprocessing to map sampled bit strings to maximal feasible solutions. To explore the prospects of quantum computational advantage, we further construct bespoke correction-in-the-loop Quantum Approximate Optimisation Algorithm instances and find that traditional Lagrange multiplier choices may not be optimal for the enumeration problem on noisy devices.

We also introduce lossless graph reductions with reconstruction rules that preserve the complete set of optimal solutions. Our results indicate that graph reduction and sampling can be effectively combined to address the enumeration bottleneck in maximum compatibility phylogeny, while retaining exact reconstruction when the residual optimum set is completely resolved.

We enhance Cherry's approach to maximum compatibility by introducing graph reduction techniques and quantum sampling.
Our classical benchmarks show a clear difference between finding one optimum and recovering the optimum set in practice. We demonstrate the utility of our methods on graphs constructed from SARS-CoV-2 sequences.
\end{abstract}

\section{Introduction}
Phylogenetic trees describe the complex evolutionary relationships of populations of organisms, providing contexts and answers to a wide range of biological questions, such as cancer research~\cite{Li2024}, drug design~\cite{Ashton2013}, and, more recently, detecting outbreaks of new SARS-CoV-2 variants~\cite{Li2020}. Constructing a truly accurate phylogenetic tree requires full knowledge of evolutionary history, something that is impossible with only present-day samples. Therefore, we must infer a “best guess” from the information that is available to us~\cite{Warnow_2017}. Since genomes of organisms are directly impacted by evolution, genome sampling provides the most informative data about evolutionary history. Advances in high-throughput sequencing~\cite{Lu2025} have led to a remarkable growth in the availability of biological data, emphasising the necessity for highly performant algorithms across the wider field, including for phylogeny. When generating phylogenetic trees one needs to pick a model which guides the choice of a metric that represents the way evolution occurs. Then, the desired output is simply the trees that are optimal in that metric.

Quantum optimisation methods provide alternative distributions over candidate solutions to hard combinatorial problems. In the present work, we use these methods as components of a modular solver rather than as a replacement for the complete classical pipeline. Our aim is to determine whether graph reduction, exact reconstruction and sampling can be combined usefully for the maximum weight clique subproblem arising in maximum compatibility phylogeny. 

We focus on the exact maximum compatibility algorithm of Cherry~\cite{Cherry2017}. A central subproblem in this algorithm is the enumeration of maximum weight cliques in a compatibility graph. This differs from the conventional optimisation task in an important way: finding one maximum weight clique is not sufficient when the downstream procedure uses a collection of equally optimal solutions to construct a strict consensus tree.

Maximum compatibility is less widely used than criteria such as maximum parsimony or maximum likelihood, but it remains relevant when some characters are not consistent with a single tree. This can occur, for example, in the presence of homoplasy or convergent mutations. Cherry's implementation is also used in pathogen phylogeny applications \cite{Cherry2017,sayers2021database}. We therefore use maximum compatibility as a concrete biological setting in which the diversity of optimal combinatorial solutions has a direct downstream role.

In particular, one scenario where maximum compatibility can be useful is in the presence of mutations that have been acquired through convergent evolution, such as the acquisition of antimicrobial resistance mutations following the release of a new drug.
These mutations do not naturally fit within a phylogenetic tree, but Cherry has shown a small percentage of such mutations is harmful to Maximum Parsimony methods and we also validated this is true for the Maximum Likelihood implementation in \emph{iqtree3}~\cite{Wong2026}

Even the simpler problem of finding a single maximum weight clique is NP-hard and has strong worst-case inapproximability results \cite{Karp1972,hastad1996clique}. However, this worst-case complexity does not by itself describe the behaviour of the graphs produced by the phylogenetic pipeline. We therefore evaluate exact and heuristic solvers on graphs derived from SARS-CoV-2 sequence data, and separately examine their ability to find one optimum and to recover distinct optima.

Our main contributions are as follows: In section \ref{sect:classical}, we compare exact and heuristic clique solvers on graphs produced by the maximum compatibility pipeline. The results show a clear distinction between finding one optimum and obtaining broad coverage of the optimum set. We introduce a deterministic repair procedure for sampled bit strings in section \ref{sect:postprocess}, and incorporate the repaired solution quality into the QAOA optimisation objective. In section \ref{sect:lagrange}, we also study a relaxed penalty regime in which the Hamiltonian is used to produce candidates for repair rather than to encode feasibility exactly. In section \ref{sect:IBM}, we evaluate the resulting sampler on a $64$-qubit quantum circuit and compare it with a random-shot baseline under the same sampling budget. We give graph reduction and reconstruction rules in section \ref{sect:qpreprocess} that preserve the complete set of maximum weight cliques. These rules are designed specifically for the enumeration problem rather than for the recovery of only one optimum. In section \ref{sect:dwavebounds}, we use a hybrid solver to seek stronger lower bounds and vertex colouring-based upper bounds during preprocessing. Our method is modular: when preprocessing produces a sufficiently small graph, the residual instance is completed classically and exactly.

\section{Preliminaries}
\subsection{Phylogeny}
Phylogenetic trees have been fundamental to understanding evolution since the very beginning of the field - Charles Darwin sketched one in his notebook soon after his trip aboard \textit{The Beagle}, where he conceived of the idea of natural selection.

A phylogenetic tree is a tree whose leaves correspond to sampled taxa and whose internal structure represents inferred evolutionary relationships. Depending on the application and the construction method, the tree may be rooted or unrooted. Molecular phylogenetic methods commonly take a multiple sequence alignment (MSA) as input. An MSA represents a hypothesis of positional homology between the input sequences; the phylogenetic procedure then uses the aligned characters to infer a tree according to a chosen criterion \cite{Warnow_2017}.

It should also be noted that there are also many methods of constructing MSAs, each representing a different philosophy. This in itself is a hugely complex process, and one that we shall isolate from the construction of the phylogenetic trees for now. Together, this yields a wide range of potential methods of constructing a phylogenetic tree from a set of sequences, all answering somewhat different questions. 

In this work, we consider the maximum compatibility criterion. The objective is to find a largest-weight collection of characters that can be realised together on one tree with the minimum possible number of state changes. This differs from maximum parsimony, another commonly used method, which minimises the total number of changes over all retained characters. However, the requirement to consider all characters can be problematic, for example in the case where there are random, noisy characters~\cite{Scotland2015}.

One of the latest maximum compatibility phylogenetic tree algorithms comes from Joshua Cherry~\cite{Cherry2017}.

\subsubsection{Cherry's Maximum Compatibility Algorithm}
Phylogeny operates in terms of ``characters'', observable traits of the sample organisms. For us, this corresponds exactly to the columns of the MSA.

The algorithm works with binary characters after constant and unsuitable columns have been removed \cite{Cherry2017}. For a variable binary character, the minimum possible number of state changes is one. In the absence of ambiguous states, pairwise compatibility of binary characters is sufficient for compatibility of the complete set.

Ambiguous bases, represented by \texttt{N} in the input considered here, make the problem more involved. Two characters may admit a pairwise compatible resolution without a single resolution being valid for every character in a larger set. The algorithm therefore resolves ambiguities iteratively and repeats the maximum weight clique calculation as required. Maximum weight clique enumeration is consequently a central
combinatorial subproblem of the complete phylogenetic procedure, rather than a complete one-step reduction of every ambiguous instance.

The characters form the vertices of a compatibility graph, with an edge between two characters when they are pairwise compatible under the current resolution. A maximum weight clique represents a maximum weight set of pairwise compatible characters, from which we obtain an optimal character set if we are able to fully resolve all ambiguities. All maximum weight cliques are retained because different optimal character sets lead to different resolved trees. The final output is obtained from the strict consensus of the resulting tree set, which retains only relationships that occur in every tree in that set.

Therefore, the main optimisation problem is to enumerate all maximum weighted clique for an input weighted graph. This is a computationally difficult challenge and even the simpler problem of finding a single maximum clique is well-known to be NP-hard~\cite{Karp1972}. There are some preprocessing refinements that are made to improve the performance of the algorithm. First, the consolidation of similar characters: consider two compatible characters that share the same pattern of compatibilities among the other characters; if one character is in the maximum clique, the other must be too. Therefore, Cherry combines these characters into a single vertex with a weight representing the number of characters that vertex represents. As a result, the optimisation problem we wish solve then becomes a slight variant of the maximum clique problem: the \textit{Maximum Weighted Clique (MWC) Enumeration Problem}. Second, the algorithm further reduces the size of the input to the MWC problem in a variety of ways. All such reductions are made to determine the vertices for which their inclusion in an MWC can be determined with certainty. 

\subsubsection{Preprocessing}
Let us first describe the preprocessing performed in~\cite{Cherry2017} in more detail, as this is a crucial part of the algorithm that has dramatic effects on the graph size, enabling very large instances to be manageable by the algorithm. It prunes vertices of the graph for which we can determine their inclusion in \emph{all} MWCs with certainty. If it can never be in any MWC, we can simply ignore it; and if it is in every MWC, we can simply store it separately and add it back at the end. As the main MWC enumeration portion of the algorithm is clearly the most inefficient part, any reduction in graph size prior to entering said subroutine provides worthwhile improvements. It should be mentioned that as the desired output is all MWCs, the pruning here is necessarily weaker than one might expect: if a vertex features in even one of the myriad of MWCs, it cannot be removed in order for said MWC to be available as a potential output for the enumeration sub-routine. This turns out to place quite a substantial limitation on the pruning, and one we address in section \ref{sect:qpreprocess}.

The key to the preprocessing part of the algorithm is the repetition until the reduction strategies can have no further impact on the size of the graph. This strategy allows us to iteratively uncover and remove vertices that cannot be in a clique without resorting to complex checking procedures.

Each loop starts with a greedy algorithm to obtain a clique that functions as an approximation to the MWC. This yields a lower bound for the MWC size, and allows the algorithm to more efficiently rule out solutions. The algorithm achieves this by systematically removing the vertex with most conflicts (in the MWC case, a ``conflict'' would be a missing edge) to the remaining subgraph until there are no conflicts left, giving us a clique. It should be noted that clearly this clique is not necessarily optimal, but is generally sufficient for the next steps. 

This is followed by a greedy vertex colouring algorithm, which takes the ordering from the previous step and assigns each vertex the lowest colour that does not conflict. Given a vertex colouring, two vertices of the same colour by definition cannot share an edge, so cannot be in the same clique. 

Suppose that the graph induced by the neighbourhood, $N_G(v)$, of a vertex $v$ has a proper colouring with colour classes $P_1,\ldots,P_k$. A clique containing $v$ contains at most one vertex from each colour class. Therefore,
\[
    U(v)
    =
    w(v)+\sum_{r=1}^{k}\max_{u\in P_r}w(u)
\]
is a valid upper bound on the weight of every clique containing $v$. Different proper colourings may give different weighted bounds even when they use the same number of colours.

It should be noted that the converse is not true: simply picking a vertex from each colour does not guarantee a clique, as having different colours does not guarantee they share an edge. Combining this upper bound with the lower bound established by exhibiting a clique, this gives a pruning algorithm specifically for the enumeration of all optima: a vertex $v$ can be removed only when the upper bound based on colouring $U(v)$ on the weight of any clique containing $v$ is strictly smaller than the current lower bound $L$. Equality cases must be retained, since a vertex with $U(v)=L$ may still belong to a maximum weight clique if the current lower bound is already optimal.

While improving the number of colours required to colour the graph does not directly cause an improvement in our bounds, it reduces the number of terms we must sum over to establish the upper bound for each vertex, leading to generally lower upper bounds. Clearly, the specifics of the colouring are also important for establishing our bounds. In the ideal world, we would find the optimal colouring for each vertex and its neighbours, which would allow us to more accurately prune the graph. However, this would dramatically increase the computational complexity of a step that is looped and is not guaranteed to produce results. Therefore, the algorithm in~\cite{Cherry2017} uses a single greedy colouring, which approximates an optimal colouring over the whole graph, and hence also over each vertex and its neighbours. This allows us to produce a more fine-tuned approximation of the MWC in a vertex's neighbour without any significant processing per vertex, only a single vertex colouring at the beginning. In comparison to the MWC problem, the vertex colouring problem is more complex due to the number of colour assignments per vertex, reducing the efficacy of the greedy algorithm, suggesting the potential for improved solution quality via quantum approaches.

From there, we can remove vertices that are more straightforwardly incompatible with any MWCs: if the weight of a vertex is greater than the sum of the weights of its neighbours, we say it is ``dominant'', and it is always better to include the vertex rather than any subset of its neighbours. Therefore, we can store the vertex to reintroduce at the end, and discard the entire neighbourhood.

When the graph is particularly large and the previous steps have been ineffective, a more aggressive reduction is used, which checks specifically for cliques within a vertex's neighbours. The sum of weights within a particular clique must be lower than the sum of all neighbours, and hence, at the (computational) cost of needing to find cliques, we can rule out vertices with complex neighbourhood graph structures that could not be previously ruled out. Noting that we do not need the MWC within the neighbours, just the existence of a clique with weight sum higher than the vertex weight, the algorithm in~\cite{Cherry2017} simplifies this step by stopping the recursive MWC as soon as it finds a clique of an appropriate weight. 

\section{Classical Solvers}\label{sect:classical}
A survey of the maximum clique problem classical solvers can be found in \cite{marino2024short} and \cite{erhardt2022engineering}.

Two performant exact clique solvers include \emph{IncMaxCLQ} \cite{li2013incmaxclq} and \emph{CliSAT} \cite{san2023clisat}. \emph{IncMaxCLQ} was state of the art when published, but it has since been surpassed by \emph{CliSAT}. We benchmark both solvers on real world graphs produced from Cherry's phylogeny tool~\cite{Cherry2017}.

We also explore \emph{FastWClq2} \cite{cai2021semi} and \emph{SCCWalk4L} \cite{wang2020sccwalk} approximate solvers, along with our own \emph{AClique} tool.
These search a graph for a fixed amount of time, unless an exact solution is determined early, and return the best clique found. They work with approximate methods using random seeds meaning repeated runs can yield different examples of the maximum clique.

This is an important point as Cherry's algorithm as initially written obtains all cliques with the maximum size and uses these to form a consensus tree. This is a standard method used to merge competing evidence into a single phylogenetic tree \cite{adams1972consensus}, \cite{bryant2017can}. However the original implementation of Cherry's code has memory usage problems as the number of equally weighted maximum cliques grows significantly as the graph grows.
Additionally even when the number of distinct cliques of maximum size is low, it may have enumerated many more maximal cliques without yet knowing they are not maximum \cite{eblen2012maximum}.
For practicality reasons we extended Cherry's tool to limit the number of stored cliques to a fixed size, with an eviction strategy for ensuring even distribution across the entire search run.

The best exact solvers are fast due to clever vertex ordering and branch-and-bound pruning, but they also yield the same instance of a maximum clique every time. It may be possible to keep backtracking and finding new cliques, but these will likely all be in the same domain of the wider search space and potentially unhelpful for consensus tree generation unless a full enumeration is tackled.

Conversely the approximate clique finders can return a different clique each execution, but usually require many execution runs. These approximate tools may be modifiable to yield many trees in a single run, but this is a complex task and considered out of scope for our evaluation purposes. Nonetheless we implemented our own approximate clique finder, \emph{AClique}, based on genetic algorithms, which can yield up to a specified number of cliques of the maximum size found in a single execution run.

\subsection{Classical Benchmarking}
The primary data set for our benchmarking, both classical and quantum, is a database of SARS-CoV-2 (COVID-19) genome sequences. This genome is clearly biologically relevant while being small (29903 base pairs). It is one of the largest studied organisms which allows us to test large cohorts.

Our test corpus contains 271,417 sequences, from which we choose a random selection varying between 50 and 20,000.
The COVID genome has very few insertions, which means we can use \emph{minimap2} to align our chosen sequences back to the canonical reference as a suitable proxy for performing a full multiple sequence alignment.
This aligned sequence is then the input to Cherry's \emph{compat} tool.
We modified \emph{compat} to output the weighted graph and an expanded unweighted equivalent graph in DIMACS format after the modified graph reduction preprocessing steps had been applied.
The graphs have a very high edge density (the percentage of possible edges that are present), with most larger graphs having $>= 99\%$ density, favouring some classical tools over others.

\subsubsection{Exact Maximum Clique Solvers}

\begin{figure}[H]
\centering
\includegraphics[scale=0.65]{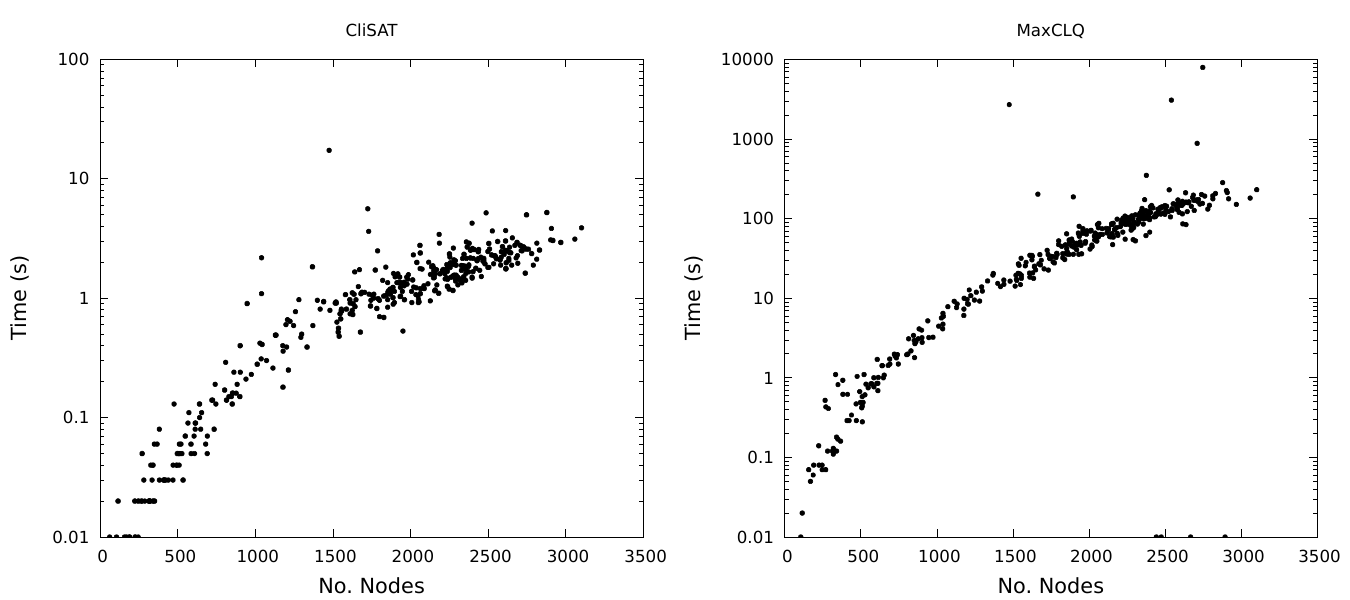}
\caption{Benchmarks of exact maximum clique solvers, \emph{IncMaxCLQ} and \emph{CliSAT}. The results derived from graphs taken from random subsets of size 100 to 20,000 sequences in increments of 50. Times are the total execution time.}\label{Fig:ExactCLQ}
\end{figure}

We benchmarked \emph{IncMaxCLQ} and \emph{CliSAT} on the unweighted graphs saved from our modified version of Cherry's \emph{compat} tool.

Figure~\ref{Fig:ExactCLQ} shows the time taken as a function of the number of nodes in the graph. Both \emph{CliSAT} and {IncMaxCLQ} show a polynomial base-line for CPU growth, which can cover the bulk of the execution runs. However there are outliers which are expected to follow an exponential CPU growth. The tools can be ran with a fixed time limit, returning the best result found within that time.

Both of these tools are feasible even for relatively large numbers of nodes, and they are both considerably faster than Cherry's implementation which ran for over a month on some of these data sets.
However Cherry's implementation is enumerating all maximum weighted cliques rather than returning a single exemplar and is thus not comparable in complexity.
We could replace the greedy lower-bound clique size step in Cherry's algorithm with \emph{CliSAT} to use the exact figure instead. However improving Cherry's exact implementation is out of the scope of this work and it is unlikely to have a significant impact given the exhaustive search is still required.

\subsubsection{Approximate Maximum Clique Solvers}

The approximate solvers run for a defined length of time, returning the best clique found within that time. They also report the time taken to find that best clique, which can help with tuning the time limit, but they do not bail out early unless they can prove there is no larger clique present.
Armed with knowledge of the actual maximum clique size (such as from a \emph{CliSAT} run), all of these approximate clique finders could be made to exit once they have found the maximum clique (or a sufficient number of them for \emph{AClique}).

We evaluated \emph{FastWCLQ2}, \emph{SCCWalk4L} and our own \emph{AClique} tool, with full results available in the Supplementary Materials. Here we focus on the best performing tool, \emph{SCCWalk4L}.

The input graphs come from the modified graph preprocessing and node reduction steps outlined above.
We colour code the results by the quality of the clique found, defined as the size of the maximum clique found from the approximate tool divided by the maximum clique as found by \emph{CliSAT} (expressed as a percentage).

\begin{figure}[H]
\centering
\includegraphics[scale=0.65]{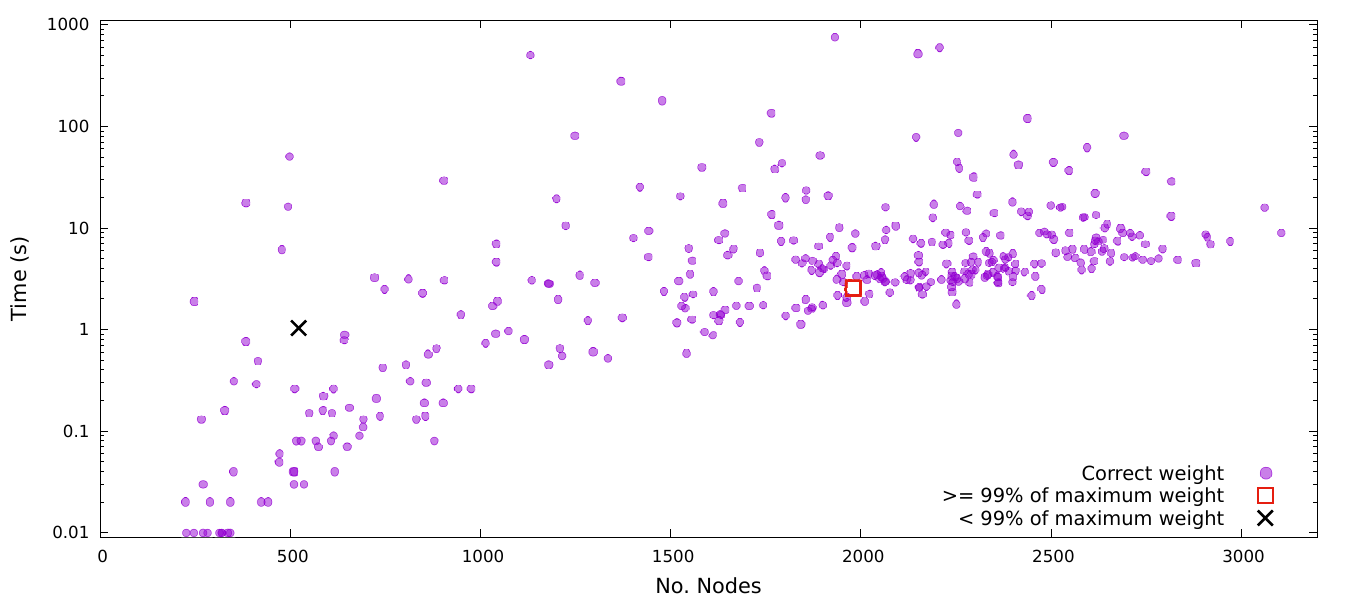}
\caption{Benchmarks of the approximate solver \emph{SCCWalk4L}. This returns a single clique after a time limit of 1000s. Colours represent the quality of result, showing perfect maximum cliques, imperfect cliques found with at least 99\% of the maximum clique size, and cliques found below 99\%. Times shown are the time to find the best solution, which is typically considerably less than the fixed specified execution time.}\label{fig:sccwalk4l-1}
\end{figure}

The results from \emph{SCCWalk4L} are shown in Figure~\ref{fig:sccwalk4l-1}.
Of 385 experiments, only 2 failed to find the maximum clique.
However the first of those failures did have the maximum clique correctly found by both \emph{FastWCLQ2} and \emph{Aclique} so there is room for a multi-algorithm approach.
None the less, this shows that some particular cliques are hard to resolve.
Remember too that we would need to run \emph{SCCWalk4L} 100s of times to build a consensus tree.

Our \emph{AClique} tool fell between \emph{FastWCLQ2} and \emph{SCCWalk4L} for quality of results, but returned many cliques in a single run.
It is likely the results could be improved further with more advanced techniques.

There is also some doubt over the randomness of results given by repeated running of the approximate MWC solvers with different random seeds.
On a test graph with 291 nodes, 10,000 runs of \emph{SCCWalk4L} found 834 distinct cliques with maximum weight, of 896 possible.
This may still be sufficient to produce the same consensus tree, but random sampling should have given us all cliques with significantly fewer trials, demonstrating a significant sampling bias.
This was not universally the case though, with another graph giving all cliques rapidly and in line with statistical expectations.
It remains the case that even some simple graphs are challenging for enumeration of maximum cliques.

\section{Quantum Formulation}
As a first step in creating an efficient quantum version of this algorithm, we start by attempting to solve this embedded combinatorial optimisation problem using quantum approaches. First, let us formalise the problem.\\

\begin{defi}[The Maximum Weighted Clique Enumeration Problem]
    Given a graph $G=(V,E)$ and a function $w:V\to\mathbb{R}$, the \textit{Maximum Weighted Clique Enumeration Problem} is to find all subsets $C\subseteq V$ that maximises $\sum_{i\in C}w(i)$ subject to the constraint that $(i,j)\in E$ for all $i,j\in C$.
\end{defi}

This differs slightly from the most common variant of this problem in two main ways: firstly, we have a weighted graph, and secondly, our desired output is all MWCs, rather than a single example.

The most common way for solving combinatorial optimisation problems on a quantum computer is using Quantum Approximate Optimisation Algorithm (QAOA)~\cite{farhi2014}. This algorithm takes a cost Hamiltonian with at most quadratic interactions and in its limit, outputs a ground state of the Hamiltonian. QAOA is an example of a solver for Quadratic Unconstrained Binary Optimisation (QUBO) problems, and we start by developing MWC solvers using this approach. In general, quantum implementations of these solvers are typically sampling-based, and hence we use this to our advantage to attempt to enumerate all optima via this sampling.

Whereas the number of edges within a clique changes with the size, $\frac{n(n-1)}2$ edges for a $n$-vertex clique, the number of pairs of vertices that does not share an edge does not change, it is always 0. Hence, rather than counting the number of edges in a potential solution which requires comparing a non-trivial expression, it is much simpler to count the number of missing edges. This inspires us to take the complement of the graph and instead find a maximum weighted subset of vertices that do not have any edges between any pair of vertices. This is the Maximum Weighted Independent Set (MWIS) Enumeration Problem.\\

\begin{defi}[The Maximum Weighted Independent Set Enumeration Problem]
    Given a graph $G=(V,E)$ and a function $w:V\to\mathbb{R}$, the \textit{Maximum Weighted Independent Set Problem} is to find all subsets $C\subseteq V$ that maximises $\sum_{i\in C}w(i)$ subject to the constraint that $(i,j)\notin E$ for all $i,j\in C$.
\end{defi}

First, let us describe the general QUBO approach. The main step involves constructing a cost function representing the specific input problem. In the broadest terms, we simply want to choose a subset of the vertices that maximises the total weight, subject to some constraints given by the edges. Since the optimisation methods we wish to employ solve unconstrained optimisation problems, we utilise a Lagrange multiplier to express the constraints. Then for a (now complement) graph $G=(V,E)$, the cost function that we want to maximise takes the form:
\begin{equation*}
    L(\mathbf{x})=\sum_{i\in V}w_ix_i-\lambda\sum_{(i,j)\in E}x_ix_j,
\end{equation*}
where the first sum is over the nodes of the complement graph and the second sum over the edges.

For some putative solution subgraph, each $x_i$ takes value 1 if vertex i is in the subgraph, and 0 otherwise, and $w_i$ is the weight of vertex i. Therefore, we only sum over the weights we include in the subgraph. In the second term, $x_ix_j$ takes the value 1 if and only if both $x_i$ and $x_j$ is 1, representing the case where the vertices on the ends of each edge are both in the subset. Then, given a sufficiently large Lagrange multiplier, any subgraph that maximises $L(\mathbf{x})$ will require none of the edges in the graph to be included, i.e. a MWIS, as they get heavily penalised. Notably, all terms are no more than quadratic, so this cost function is naturally in the form that the QUBO method aims to solve.

The QAOA approach requires a slight modification over the general QUBO method, as we must have our binary variable take values over $\{-1,+1\}$, rather than $\{0,1\}$. Therefore, we require a further step to convert the cost function to a Hamiltonian. The general form of a cost Hamiltonian is
\begin{equation*}
    H_C=\sum_{ij}Q_{ij}Z_iZ_j+\sum_ib_iZ_i,
\end{equation*}
where the sums are over the qubits and the $Z_i$ is a Pauli-Z gate on the $i$-th qubit.

In order to rewrite our cost function in this form, first we transform the binary variables $x_i\in\{0,1\}$ to $z_i\in\{-1,1\}$ via 
\begin{equation*}
    x_i=\frac{1-z_i}{2}\quad\text{or}\quad z_i=1-2x_i.
\end{equation*}
Then our cost function becomes:
\begin{align*}
    L(\mathbf{x})&=\sum_{i\in V}w_i\frac{1-z_i}{2}-\lambda\sum_{(i,j)\in E}\frac{1-z_i}{2}\frac{1-z_j}{2}\\
    &=\sum_{i\in V}w_i\frac{1-z_i}{2}-\lambda\sum_{(i,j)\in E}\frac{(1-z_i)(1-z_j)}{4}\\
    &=\sum_{i\in V}\frac{w_i}{2}-\sum_{(i,j)\in E}\frac{\lambda}{4}-\sum_{i\in V}\frac{w_iz_i}{2}+\frac{\lambda}{4}\sum_{(i,j)\in E}(z_i+z_j)
    -\frac{\lambda}{4}\sum_{(i,j)\in E}z_iz_j.
\end{align*}
Now, we can note that for the $\sum_{(i,j)\in E}(z_i+z_j)$ term, the number of times each $z_i$ is counted is exactly the degree of the vertex, so we can simplify this term to $\sum_{i\in V}\text{deg}(i)$.
Furthermore, there are several things that do not affect the argument that maximises the function: constant terms and multiplicative factors. Therefore, we can remove the first two terms and multiply the whole expression by 4. This gives an equivalent maximization objective, up to additive and positive multiplicative constants
\begin{equation*}
    L(\mathbf{x})=\sum_{i\in V}\left(\lambda\cdot\text{deg}(i)-2w_i\right)z_i
    -\lambda\sum_{(i,j)\in E}z_iz_j.
\end{equation*}
In most QUBO solvers, including all of the ones we consider here, the solver actually minimises the loss function or Hamiltonian, so in our implementation we implement the negative of the above functions. We implement these methods on leading quantum platforms such as D-Wave and Qiskit for IBMQ in order to validate our formulations and evaluate their performance. We use the constructed solvers on small instances of real SARS-CoV-2 genomes.

A crucial factor for demonstrating quantum advantage is the ability to scale up to larger instances, beyond what classical approaches could handle. Luckily, the qubit usage scales 1-to-1 with the number of vertices in the reduced graph. For the QAOA approach, we also need to consider the depth and the gate count, which also scales approximately linearly with the size of the reduced graph:

\begin{figure}[H]\centering
\subfloat[Qubit Count]{\label{a}\includegraphics[width=.45\linewidth]{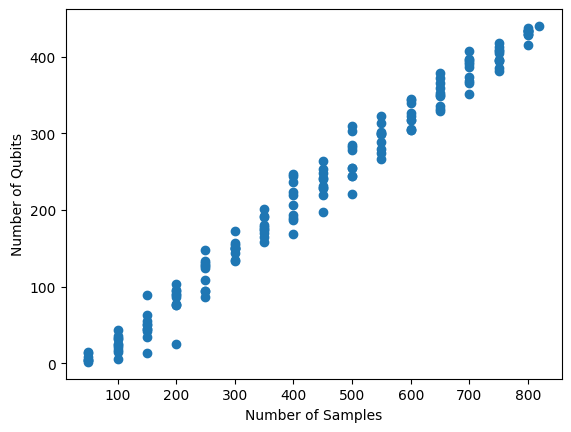}}\par
\subfloat[Depth]{\label{b}\includegraphics[width=.45\linewidth]{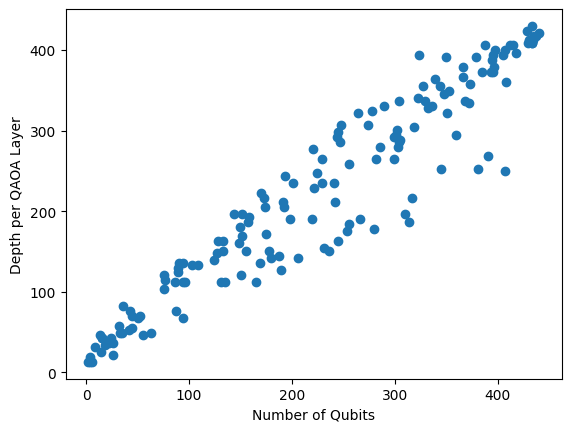}}\hfill
\subfloat[Gate Count]{\label{c}\includegraphics[width=.45\linewidth]{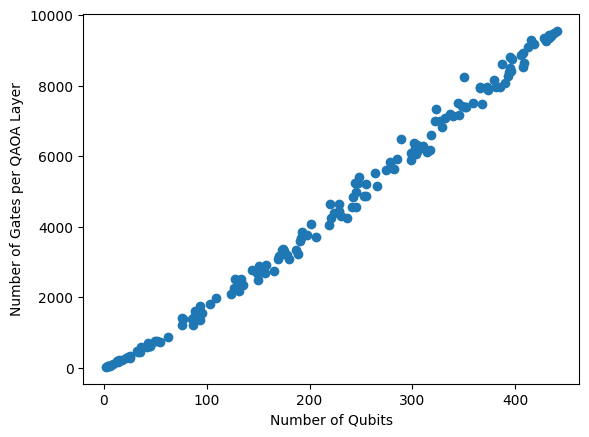}}
\caption{Scatter plots of the depth and gate count against number of qubits of the QAOA circuits required to solve the MWC problem on various subsets of SARS-CoV-2 genome samples.}

\end{figure}

An important aspect of the problem to keep in mind is the requirement to find \textbf{all} of the MWCs within the graph. Measuring the QAOA circuit in the computational basis yields a single bitstring that represents a subgraph. Therefore, we approach this enumeration by repeatedly measuring executions of the same circuit. This is a well-established protocol in quantum computing known as taking ``shots''. Commonly, this process is just used to find the ground state energy of the Hamiltonian, but for us, we are interested in the states that give this lowest energy. Therefore, we hope to find all of the MWCs by running shots and keeping track of the states that produce these lowest energies. An important question of this approach is how many shots are required to find all of these cliques? In the case that any MWC is equally likely, this is exactly the well-known \textit{Coupon Collection Problem}, where a coupon collector is wanting to collect all $N$ coupons by drawing with replacement. The expected number of trials for this problem is $\Theta(N\log N)$.

\subsection{Postprocessing}\label{sect:postprocess}
In this approach, we no longer strictly enforce the edge constraints. Therefore, in sampling low energy solutions, while we hope to mostly obtain ground states that correspond to the MWCs, we may still get results that are not optimal. They may contain edges of the complement graph, or simply just not be of maximum weight.

For the former, this is easy to check: simply construct the subgraph represented by the measured bit-string, and check for any edges. Therefore, we have an efficient path towards retrieving at least a feasible solution. In the latter scenario, given a feasible solution that is not optimal, a natural question to ask is whether we can efficiently reach an optimal solution from it. The answer is no, as a sub-optimal solution need not be contained with in an optimal one, and hence does not necessarily give a natural path to one. 

One can consider an outputted result as a graph in its own right, albeit a graph that one would hope is closer to an optimal solution than the original input graph. Another (classical) method that can take a graph and output a more optimal solution is the conventional greedy algorithm for the clique problem. Recall that the algorithm works by removing vertices of the lowest degree (i.e. missing edges to the most vertices) until the remaining set of vertices has full connectivity. This algorithm is guaranteed to run in polynomial time returns a feasible solution (which is unfortunately is not guaranteed to be optimal, but is at least maximal). We can combine these two methods and consider the classical greedy algorithm as a postprocessing method, in order to ensure that the subgraphs that we consider are at least cliques, satisfying all edge constraints. Furthermore, given a clique, we can check for any additional vertices connected to all the vertices in the clique, and hence the union of the two is still a clique. Repeating this until we no longer add any more vertices gives us a maximal clique, that is, a clique that is not contained within a larger one.

By doing these two steps to all results from our quantum sampling-based algorithms, we can at least insist the subgraph we consider is a maximal clique, a much smaller set of graphs (there are at most $3^{n/3}$ of them) than arbitrary subgraphs of the input graph.

A crucial step within the QAOA algorithm is the classical optimisation loop to find optimal parameter values based on the energy of each set of samples. However, as we have established in the section above, changing the postprocessing step, which enforces the edge constraints has a dramatic effect on our solutions, and their energies. Therefore, it would be remiss of us not to include this efficient classical postprocessing within the classical optimisation loop, in order for it to optimise over the true figure of merit, rather than some intermediate value.

\subsection{The QUBO Lagrange Multiplier}\label{sect:lagrange}
Conventional wisdom suggests that the Lagrange multiplier used to turn a constrained problem into an unconstrained problem must be large enough such that the ground state corresponds to the case where all constraints must be satisfied. For the MWC problem, this corresponds to the Lagrange multiplier being at least as large as the largest vertex weight. For example, a common choice is twice the largest vertex weight.

However, given the above postprocessing, we can efficiently ensure our final results must satisfy our edge constraints and hence it is less essential that our quantum sampling-based solver strictly enforces these constraints. Indeed, while the ground states of the cost Hamiltonian with large Lagrange multipliers correspond to exactly the MWCs of the graph, the states with energies that are low but not optimal do not necessarily correspond to states that are very close to any of the MWCs as one could have a clique with weight that is just 1 off the maximum, but has a completely different graph structure.

We find that by taking a Lagrange multiplier that is smaller, the ground states no longer correspond to the MWCs, as adding a vertex with a few missing edges to the rest of the putative clique can still decrease the Hamiltonian energy. Taking small graph examples for which we can enumerate all possible subgraphs, and by applying the postprocessing methodology above to the low-energy subgraphs, we see that taking the ratio between the Lagrange multiplier and the highest vertex weight to be 0.1 gives the highest proportion of subgraphs that yield MWCs.

\begin{figure}[H]\centering
\subfloat[20 Vertices]{\includegraphics[width=.45\linewidth]{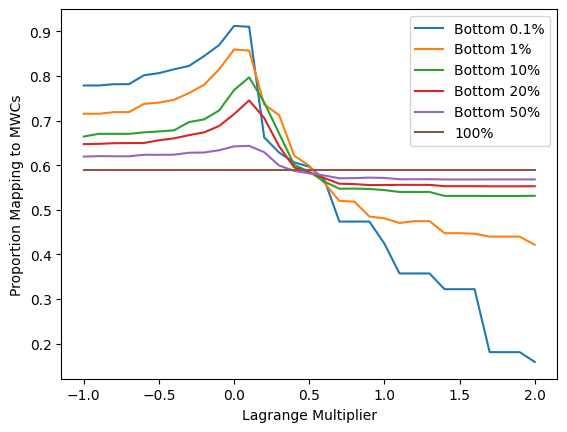}}\hfill
\subfloat[23 Vertices]{\includegraphics[width=.45\linewidth]{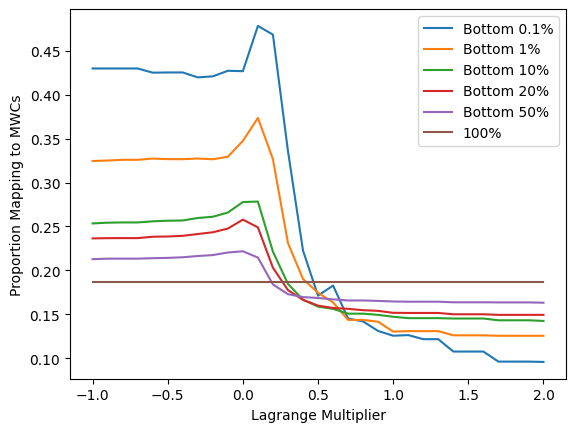}}
\caption{For different Lagrange multipliers, the low energy solutions correspond to different subgraphs. Taking different fractions of the lowest energies, we show what proportions of the corresponding subgraphs correspond to MWCs given our postprocessing algorithm.}
\end{figure}

\subsection{Hardware proof of Concept}\label{sect:IBM}
An example of a current quantum device that the general public can have access to now is the series of devices by IBM. They have up to 156 qubits. Therefore, given the 1-to-1 variable-to-qubit mapping of our QUBO formulation of the MWC enumeration problem, on an IBM devices, we can only run instances of the problem which results in a final reduced graph of 156 vertices or fewer. In the next section, we will describe our efforts to further the classical preprocessing such that larger raw input sizes results in graphs small enough to implement on these devices, but for now, let us examine the current performance of these devices.

\begin{figure}[H]
\centering
\includegraphics[scale=0.65]{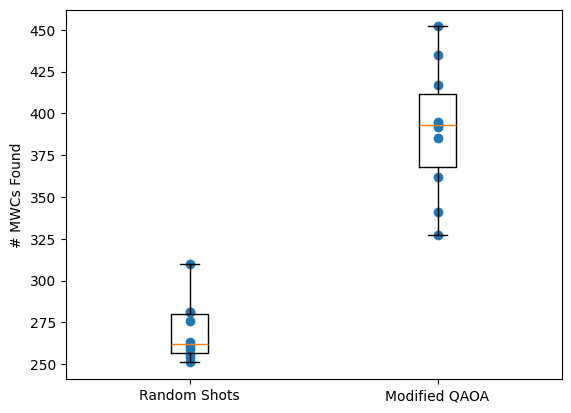}
\caption{Resulting number of MWCs found from 10 runs of $20{,}000$ shots of random shots and from our modified QAOA}.
\end{figure}

Running on \textit{IBM Heron} r3 machines (\textit{Boston} and \textit{Aachen}), we train and run our modified QAOA circuits, requiring the same 64 qubits and ~3500 two-qubit gates. Due to the noisy nature of current machines, we see a very high variance in the results from QAOA runs, but nonetheless we can see that it can clearly outperforms taking random shots, which in itself is surprisingly much faster than using an approximate classical algorithm that can produce a single MWC to yield a large set of MWCs. Crucially, the $20{,}000$ shots we choose to use based on the resources we have available is very small in the scheme of quantum experiments and allows for the massive potential to scale up the final sampling run in order to obtain a much larger set of MWCs.

The random shot baseline draws each bit independently with probability $0.5$ and applies the same postprocessing as for the QAOA samples. By sampling in this manner we uniformly sample from the space of all subgraphs.

\section{Quantum Focussed Preprocessing} 
In light of the limited capacity of quantum devices, the preprocessing of the input data to yield the smallest possible graph becomes especially important. Here, we look at extending the preprocessing in~\cite{Cherry2017} to take into account reductions that isolate the difficult core of hard problem instances at the cost of extra processing and also utilise sophisticated optimisers.

\subsection{A lossless graph ``compression''}\label{sect:qpreprocess}
The algorithm in~\cite{Cherry2017} has a preprocessing step that removes the vertices for which we can be certain on their inclusion in any putative MWCs. This reduces the graph size in which we need to find the MWCs, and the process of recovering the MWCs of the original graph is extremely simple: for each MWC of the reduced graph, simply take the set union of the MWC and the vertices we had established must be in any MWC of the original graph. Note that this process is 1-to-1 between the MWCs of the original and reduced graphs - we are simply adding the same vertices to every MWC of the reduced graph. Therefore, for our quantum QUBO/QAOA based approaches where we hope to extract all via MWCs via exploiting the use of shots, this does not cause a reduction in the number of shots required to solve the problem (for a fixed accuracy), which is one of the major resource costs. That is not to say the reduction is useless: simply by reducing the graph size, this reduces the size of the resulting QAOA circuit and hence will be cheaper, more accurate, and faster. Therefore, we looked towards reductions that could also reduce the number of MWCs within the new reduced graph, but for which we can still easily recover the full set of the MWCs for the original graph.

Here, let us work in the complement graph as it is much sparser and the native object the QAOA approach considers. Further, recall that the Maximum Weighted Clique enumeration problem becomes the Maximum Weighted Independent Set enumeration problem in the complement graph. Crucially, the solutions for the two problems are exactly the same. 

Suppose we wish to find all MWISs within a graph $G$. Call the set of MWISs $C$. The idea of these reductions is to produce a graph $G'$ such that its set of MWISs, $C'$, is easier to compute, and given $C'$, we can easily recover $C$. Let $\omega_G(v)$ be the weight of a vertex $v$ in the graph $G$.

In the most simple case, we can remove all vertices of degree 1. Let $v$ be a vertex of degree 1 that is connected to the vertex $w$ on $G$. We know that any MWIS that contains $w$ cannot contain $v$ by definition, but crucially, also that any MWIS that doesn't contain $w$ will always contain $v$ as $v$'s only conflict is with $w$, and including $v$ will increase the weight. Therefore, all cliques from the graph must contain exactly one of $v,w$, and $v$'s inclusion in an MWIS is fully determined simply by whether $w$ is in it. This is also true of the weight of $v$'s inclusion in the weight of the putative MWIS. So by considering $v$ as being included in the MWIS \textit{unless} $w$ is in the MWIS, we can consider the action of adding a vertex $w'$ (which has the same connectivity as $w$ in $G$, with the exception of the edge to $v$) to an independent set in $G'$ as adding $w$ and removing $v$ to an independent set in $G$. Therefore, we would want $\omega_{G'}(w')=\omega_G(w)-\omega_G(v)$, and the maximum weight having an additive constant of $\omega_G(v)$, which we can store separately.

This reduces the number of vertices and number of edges both by 1. But now let us consider the new vertex $w'$. It has a weight of $\omega_G(w)-\omega_G(v)$, which is no longer guaranteed to be positive. Indeed, if the weight is negative, it is never in any MWIS and so can be excluded from the graph (and so $v$ is always in all MWIS of $G$). If the weight is 0, then whether $w'$ is included in a potential IS does not change its weight, and so its inclusion is determined purely by whether its conflicts are in the MWIS. In the case where there exists a MWIS containing $w'$, there also exists a MWIS without $w'$. Therefore, we can find the MWISs of the graph without $w'$ and check whether any of the neighbours is included to decide whether there is also a MWIS containing $w'$. So when $w'$ has weight less than or equal to 0, this reduces the graph size by a further vertex and potentially multiple edges. Crucially, in doing this reduction, we may also expose further vertices, which allows us to continue to reduce the graph.\\

\begin{figure}[H]
    \centering
    \includegraphics[width=0.8\linewidth]{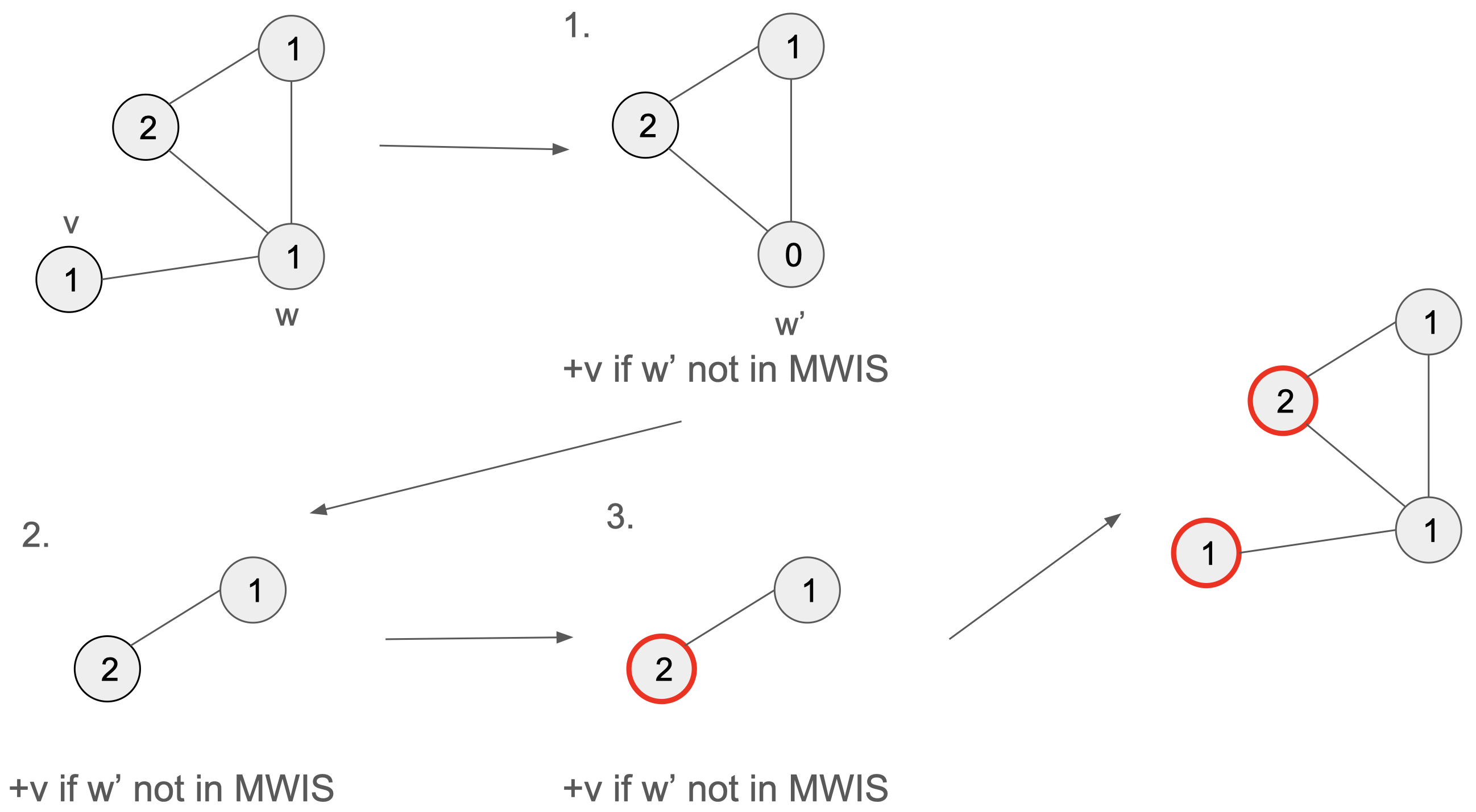}
    \caption{An example of the effects of the reduction process on the MWIS algorithm on a simple complement graph. The target vertex is $v$, which is connected to $w$. In step 1, we remove $v$ and decrease the weight of $w'$. In step 2, as $w'$ has weight 0, we remove it to deal with in postprocessing. In this case, as it is connected to both the remaining vertices, it cannot be in any MWIS. In step 3, we find the MWIS in the reduced graph, which is simply the vertex of weight 2. This is then combined with $v$ which we had removed in the beginning to give the sole MWIS for this complement graph.}
    \label{fig:ex}
\end{figure}

This concept can be extended further: consider a vertex of degree 2, $v$, connected to a pair of vertices that are also connected to each other, $w_1,w_2$. Then, we can apply similar principle to the degree 1 case. The vertex $v$ can only be in a MWIS if neither $w_1$ nor $w_2$ is in it, and crucially as $w_1,w_2$ conflicts, only one of them can be in a putative MWIS at a time. As a result, in this case we case remove $v$ from the graph and dropping the weight of both $w_1',w_2'$ by $\omega_G(v)$, knowing that as only at most one of the vertices can be in the MWIS and so we would never want to remove $v$ twice.

These two methods form part of a larger series of reductions, which are sometimes collectively known as ``kernelisation routines''~\cite{grossmann2024}. We limit our attention here to the two described above due to the complexity of verifying the reduction requirements for more complex reductions, and their limited expected utility. As mentioned above, each of these reductions will expose new vertices that may newly satisfy the conditions for reduction, whether it be Cherry's original reduction or the new ones we have developed. Therefore, instead of having these reductions as a stand-alone subroutine embedded into the algorithm, we embed it directly into Cherry's preprocessing step. This allows us to exploit Cherry's reductions and re-use them all until no further improvements in graph size via these reductions can be achieved.

Carrying out these reductions on the SARS-CoV-2 samples we have previously discussed, we see an approximately 20\% reduction in both the number of vertices and edges simply for the degree 1 reductions. Including the degree 2 reductions drops the graph sizes further, to around only 60\% of the original number of vertices and edges. 
\begin{figure}[H]\centering
\subfloat[Vertices]{\includegraphics[width=.45\linewidth]{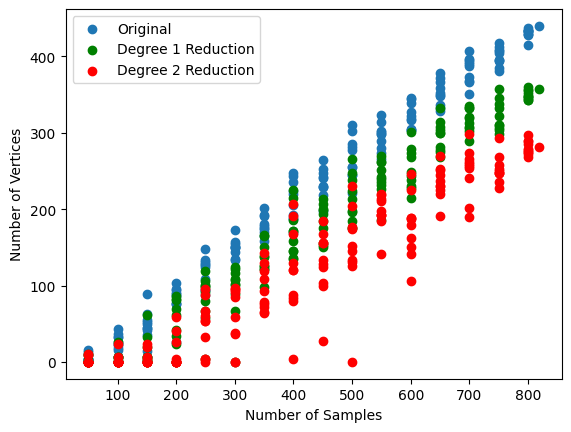}}\hfill
\subfloat[Edges]{\includegraphics[width=.45\linewidth]{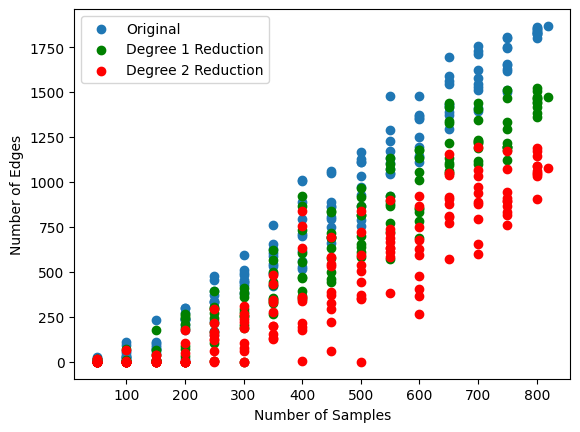}}
\caption{Numbers of vertices and conflict edges before and after the degree one and degree two reductions on the SARS-CoV-2-derived instances. All reductions retain reconstruction records for the complete optimum
family.}\label{fig:deterministic-reductions}
\end{figure}

For the instance constructed from $819$ SARS-CoV-2 samples, the original complement graph contains $440$ vertices and $1868$ edges. The degree one rules reduce this to $357$ vertices and $1473$ edges. Including the degree two rules gives a residual graph with $281$ vertices and $1076$ edges.

Thus, relative to the original residual graph, the combined deterministic reductions remove $36.1\%$ of the logical variables and $42.4\%$ of the
conflict edges. The first quantity is the corresponding reduction in the logical qubit count of the direct encoding. At fixed QAOA depth, the second quantity is the corresponding reduction in the number of logical two-qubit cost interactions in each cost layer.

\begin{table}[H]
\begin{tabular}{|c|cc|cc|}
\hline
\multirow{2}{*}{}  & \multicolumn{2}{c|}{Vertices}      & \multicolumn{2}{c|}{Edges}   \\
                   & Number & Percentage of Original & Number & Percentage of Original \\ \hline
Original           & 440    & -                      & 1868   & -                      \\
Degree 1 Reduction & 357    & 81.1\%                 & 1473   & 78.9\%                 \\
Degree 2 Reduction & 281    & 63.9\%                 & 1076   & 57.6\%                 \\ \hline
\end{tabular}
\caption{Effect of the deterministic reductions on the complement graph for the instance constructed from $819$ SARS-CoV-2 samples. The percentage columns give the residual size relative to the graph entering the new reduction stage.}\label{tab:reductions}
\end{table}
The degree one and degree two rows in Table~\ref{tab:reductions} are cumulative: the degree two result includes all reductions already made by the degree one rules.

\subsection{Improving the classical pruning}\label{sect:dwavebounds}
The original preprocessing pipeline in~\cite{Cherry2017} includes two methods to bound the resulting MWC size, and as a result eliminate vertices. We establish a minimum MWC size simply by exhibiting an example of a clique, and a vertex colouring gives an upper bound on the best clique that can contain each vertex.

The correctness of the pruning stage does not require either sampled object to be optimal. It requires only that the lower bound solution is a valid clique and that the upper bound colouring is proper. At present, we derive both the minimum clique size and vertex colouring based on simply (polynomial-time) greedy algorithms. 

As such, there are instances for which these approximations will necessarily far from optimal~\cite{hastad1996clique,Grimmett1975}. Therefore, we seek to improve on these approximation using the hybrid quantum-classical sampler, \emph{LeapHybridSampler}, on the D-Wave platform, which while only partially quantum, allows us to tackle much larger problems sizes that purely quantum platforms would allow. Crucially, in switching to a D-Wave solver for these problems rather than the greedy algorithm on a fixed, predetermined ordering based on the number of conflicts each vertex has, we allow ourselves to obtain any (hopefully closer-to-optimal) vertex colouring.

We verify and correct each candidate classically before using it. For both tasks, this is known to be possible in polynomial time. For a candidate clique, we check every pair of selected vertices, and simply greedily remove any conflicts. For a candidate colouring, we check that the endpoints of every edge have different colours, and change one endpoint to a brand new colour in the case of any conflicts. Furthermore, we retain the original classical greedy examples. Thus, stochastic sampling can affect the strength of the reduction but not its correctness, and at worst, we recover the performance of the original classical preprocessing (albeit with a longer run time).

\begin{figure}[H]\centering
\subfloat[Vertices]{\includegraphics[width=.45\linewidth]{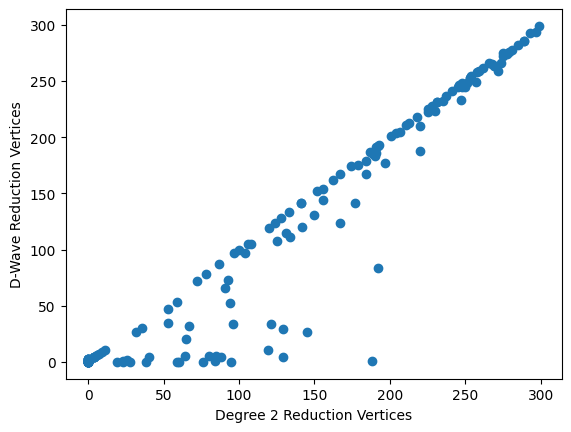}}\hfill
\subfloat[Edges]{\includegraphics[width=.45\linewidth]{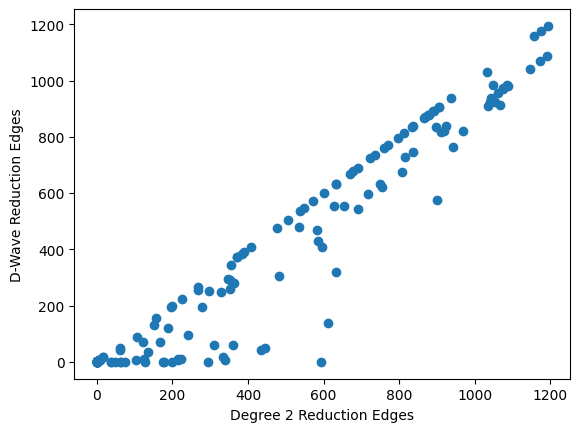}}
\caption{The reduction of number of graph vertices and edges for SARS-CoV-2 samples between the degree 2 and D-Wave reductions.}
\end{figure}

The reduction is highly dependent on the specific vertex colouring, rather than just the number of colours required. Therefore, the efficacy of the hybrid reduction is no longer deterministic and will fluctuate with the exact colourings obtained. Indeed, as the instance sizes grows, the number of different colourings giving the same weight often grows very quickly too, decreasing the chance of finding one that will yield a powerful reduction. Nonetheless, even with sampling 3 different colourings, we see a impact in our problem sizes. Furthermore, recall the purpose of this reduction: it is to remove vertices that we do not belong in any putative clique. Therefore, extra effort here does not lead to increased postprocessing, and is worth it in order to reduce the computational requirements of the much-harder problem of enumerating all cliques on the resulting graph. In a similar vein to our overall argument, we do not necessarily see this preprocessing to be useful for every input, but simply, where this reduction is effective, it is very fruitful and will allow us to tackle much larger problem sizes than before.

For our complete hybrid classical-quantum preprocessing pipeline, we see the following reductions in problem sizes:
\begin{figure}[H]\centering
\subfloat[Vertices]{\includegraphics[width=.45\linewidth]{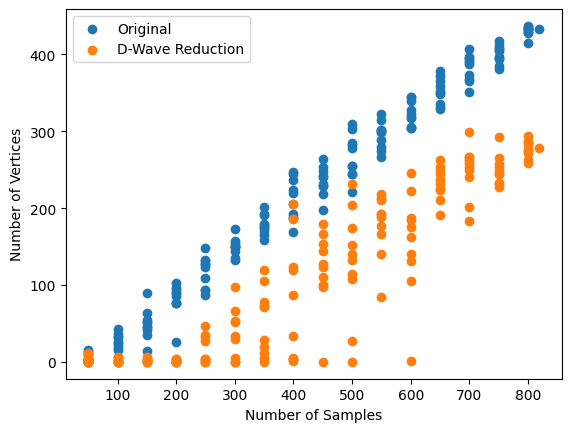}}\hfill
\subfloat[Edges]{\includegraphics[width=.45\linewidth]{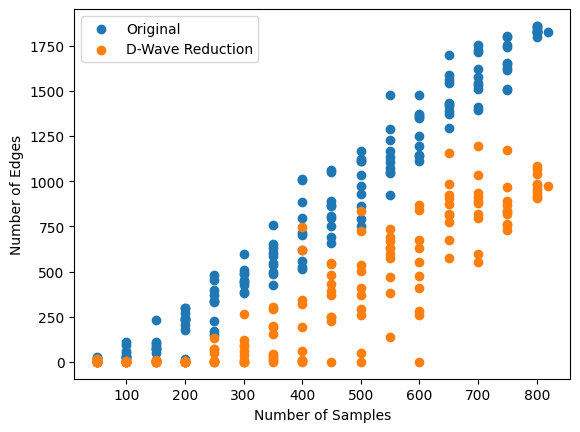}}
\caption{ The reduction of number of graph vertices and edges for SARS-CoV-2 samples between the original postprocessing and D-Wave reductions.}
\end{figure}

This reduction shrinks the graph to a much smaller instance. To illustrate its utility, the set of 350 samples was reduced to 64 vertices entirely classically, we can now reduce this to simply a 5-qubit complete (complement) graph:
\begin{figure}[H]\centering
\subfloat[Original complement graph]{\includegraphics[width=.30\linewidth]{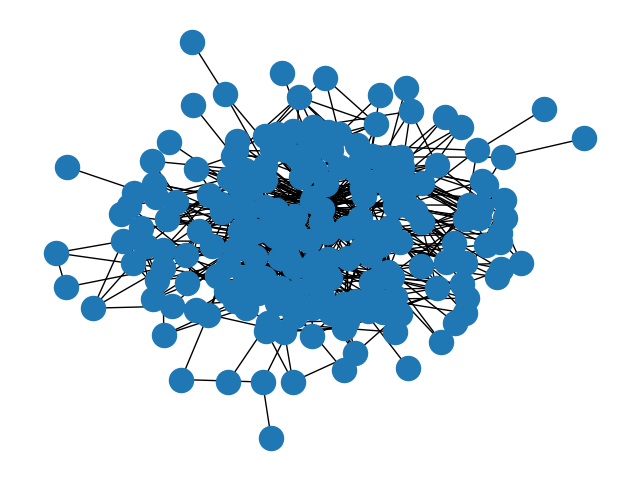}}\hfill
\subfloat[Complement graph with classical reductions]{\includegraphics[width=.30\linewidth]{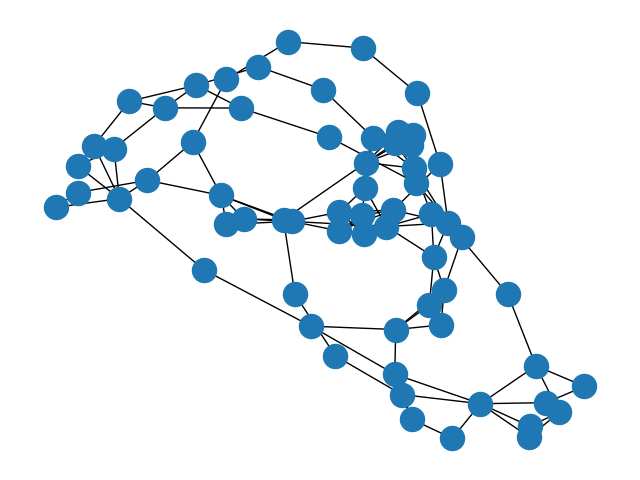}}\hfill
\subfloat[Complement graph with D-Wave reductions]{\includegraphics[width=.30\linewidth]{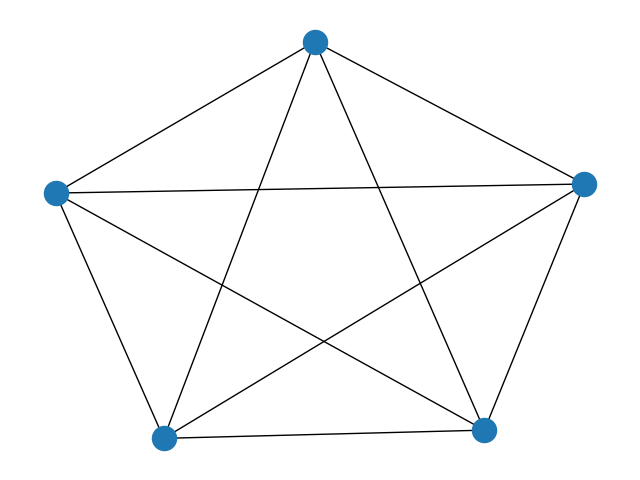}}
\caption{The graph reductions on an example set of 350 SARS-CoV-2 samples.}
\label{350red2}
\end{figure}

Now on this fully reduced graph, it is clear what the MWISs should be: simply the individual vertices of the graph that has the maximum vertex weight. In this case, they all have weight 1, so we have exactly 5 MWIS for this reduced graph, and with our MWIS reconstruction postprocessing, we can deterministically recover the \emph{entire} set of MWIS. It should be noted that we do not mandate that we must use quantum devices in every step of our pipeline. Indeed, given a very small reduced graph, it clearly does not make sense to do so. Therefore, with our quantum pipeline, we recover the potential to determine the full set of MWISs and output the \emph{exact} maximum compatibility phylogenetic tree.

It is important to stress the variability of the reductive power of our preprocessing based on the D-Wave clique and vertex-colouring samplers. Due to our limited quantum resources, we do not attempt to fully optimise the graph reductions for every instance, but simply perform a single pass of the algorithm on the instances to gain an insight on its behaviour and potential.

\section{Discussion and Outlook}
We have studied the maximum weight clique subproblem arising in maximum compatibility phylogenetic tree construction algorithm~\cite{Cherry2017}, aiming to improve the recovery of multiple optimum solutions. The main algorithmic distinction is between finding one maximum weight clique and recovering a useful portion of the complete optimum set. The numerical experiments show that these can be very different tasks: \emph{SCCWalk4L} found an optimum on $383$ of the $385$ tested instances, but repeated runs on one graph still missed $62$ of the $896$ optima after $10{,}000$ executions. A solver can therefore perform well under the usual best weight metric while providing limited coverage for applications that require several equally optimal solutions, including strict consensus tree construction.

Our contributions are twofold. The first is a more efficient reduction method that allows us to use enumeration methods that also include quantum algorithms on current devices. We apply a stricter reduction scheme by removing vertices from the complement graph and separately record the information required to reconstruct the corresponding maximum weight independent sets. This reduces the graph which is passed to the main enumeration algorithm while making the reconstruction step more computationally costly.

Remarkably, the reconstruction rules preserve every optimum. On the instance constructed from $819$ samples, the reductions decrease the graph from $440$ to $281$ vertices and from $1868$ to $1076$ conflict edges. Thus one can remove $36.1\%$ of the logical variables and $42.4\%$ of the conflict edges. These reductions benefit both classical and quantum solvers.

Our solver-agnostic approach has the potential to completely reduce the input graph: in one instance, the residual graph could be completed exactly after a reduction from $64$ to five edge-free vertices. Then the enumeration problem becomes trivial and the full optimum set on the original graph can be easily recovered.

The second contribution is a QUBO formulation of the MWC problem through the corresponding MWIS problem on the complement graph, together with modifications to the standard QAOA procedure aimed at sampling multiple optimum solutions on current devices. We use a postprocessing procedure inspired by the classical greedy algorithm for the clique problem. This procedure runs in polynomial time, removes conflicts, and extends the resulting feasible solution until it is maximal. It maps sampled subgraphs to maximal cliques, which form a substantially smaller set than arbitrary subgraphs. The procedure does not by itself establish that the result has maximum weight. A repaired clique is accepted as an MWC only when its weight agrees with a current optimum value.

Since the postprocessing procedure enforces the edge constraints, the cost Hamiltonian need not enforce them with the same strength. On the small graphs for which all subgraphs could be enumerated, we found that a smaller Lagrange multiplier produced a larger proportion of low energy states that repaired to MWCs among the tested values. In this regime, the MWCs need no longer correspond exactly to the ground states of the Hamiltonian. However, low energy states may contain conflict edges that can be removed during postprocessing, rather than forming a lower weight clique whose structure differs substantially from every MWC. This motivates the use of the Hamiltonian as a sampling objective together with deterministic repair, rather than as an exact encoding in this parameter regime.

Using a sampling method to enumerate $N$ optimum solutions requires at least $N$ samples. Under uniform independent sampling, the coupon collection model gives an expected sample count of order $N\log N$. The classical experiments show that the distribution over the optimum solutions can be strongly nonuniform, so this estimate should be treated as an idealised reference.

On a graph with $64$ vertices, the modified QAOA procedure recovered more distinct certified MWCs than the random shot baseline in each of the ten reported runs. This result demonstrates improved sample coverage on the tested instance, but it is not a claim of quantum advantage. Taken together, the results support a modular approach in which exact preprocessing, classical completion, and sampling methods are selected according to the residual graph and the required level of completeness.

\section{Code Availability}
AClique is available from \url{https://github.com/q4bio-qpg/aclique}.

The modified Cherry's algorithm is available from \url{https://github.com/q4bio-qpg/qcompat}.

\section{Acknowledgements}
This work is supported by Wellcome Leap as part of the Q4Bio Program. OY is supported by the Martingale Foundation. JB was additionally funded by the Wellcome Trust [Grant number 220540/Z/20/A]. SS acknowledges support from the Royal Society University Research Fellowship.

\bibliographystyle{plain}
\bibliography{lit.bib}
\newpage
\appendix
\section{Approximate Classical Solvers}
We chart the time taken to find the first occurrence of the best clique found, with runs of fixed times for 10, 100 and 1000 seconds per sequence subset. This time may be considerably less than the specified timeout, but the program still runs to the full timeout while looking for a better result. Thus the graphs can be seen as an indication of the typical timeout needed to be confident the correct result has been found. (This is most obvious with the later SCCWalk4L results.) The input graphs come from the modified graph preprocessing and node reduction steps outlined above. We colour code the results by the quality of the clique found, defined as the size of the maximum clique found divided by the actual maximum clique as found by \emph{CliSAT} (expressed as a percentage).

\begin{figure}[H]
\centering
\includegraphics[scale=0.65]{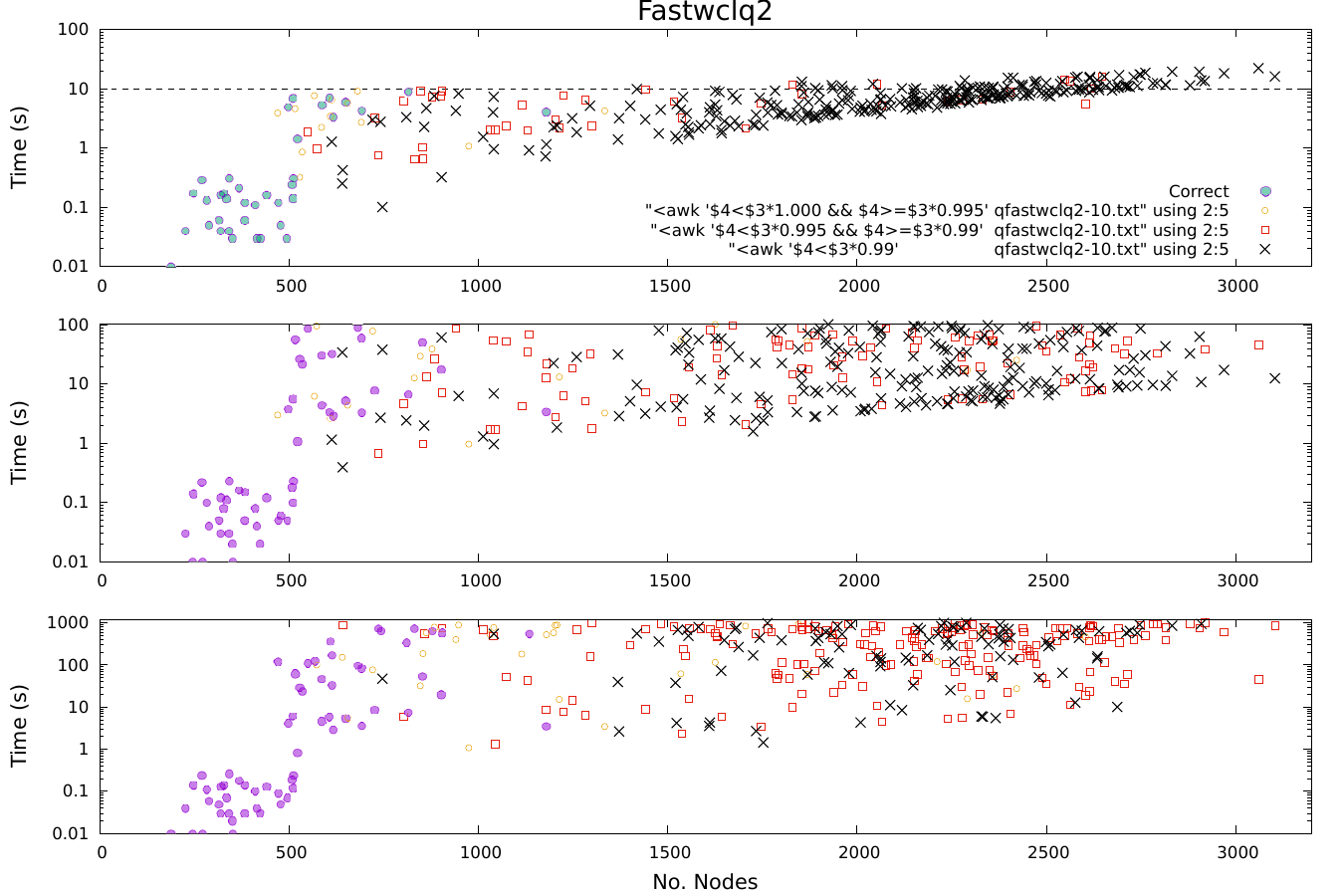}
\caption{Benchmarks of the approximate solver \emph{FastWCLQ2}. This returns a single clique after a given maximum time limit. We chart the results for 10, 100 and 1000 seconds of time. Colours represent the quality of result, showing perfect maximum cliques, imperfect cliques found of size $\geq 99.5\%$, cliques found of size between 99\% and 99.5\%, and cliques found below 99\%. Times shown are the time to find the best solution, which is typically considerably less than the specified execution time.}\label{fig:fastwclq2}
\end{figure}

\emph{FastWCLQ2} results are shown in Figure~\ref{fig:fastwclq2}.
Despite being reviewed favourably by others, it appears to dislike our dense graphs and gives poor results even with the maximum time of 1000 seconds. In the 10 second runs it also took noticeably longer than the maximum time limit. We assume this is a constant time component loading and preprocessing the graphs as it was not significant in the longer trials.

\begin{figure}[H]
\centering
\includegraphics[scale=0.65]{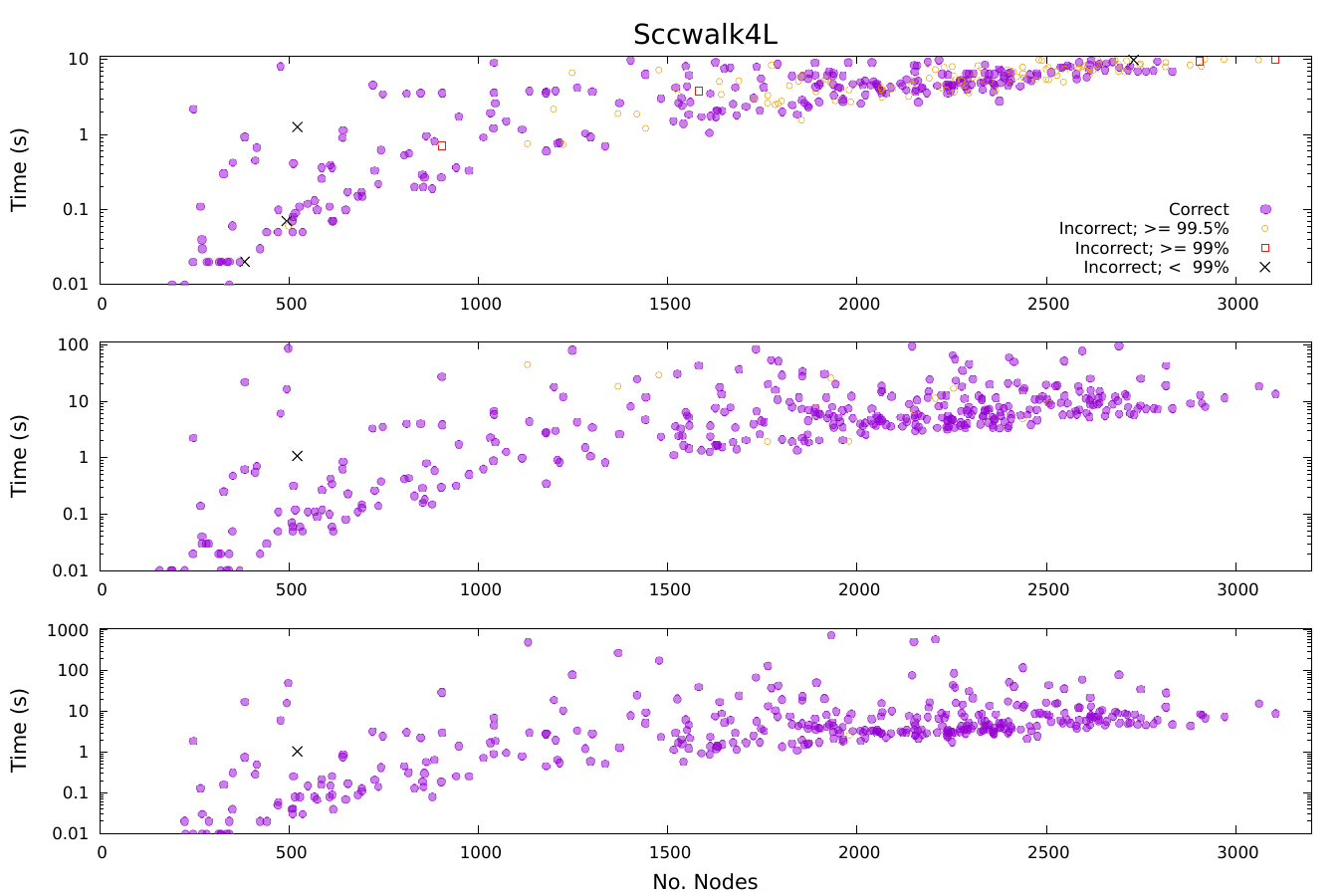}
\caption{Benchmarks of the approximate solver \emph{SCCWalk4L}. This returns a single clique after a given maximum time limit. We chart the results for 10, 100 and 1000 seconds of time. Colours represent the quality of result, showing perfect maximum cliques, imperfect cliques found of size $\geq 99.5\%$, cliques found of size between 99\% and 99.5\%, and cliques found below 99\%. Times shown are the time to find the best solution, which is typically considerably less than the specified execution time.}\label{fig:sccwalk4l}
\end{figure}

\emph{SCCWalk4L} results (Figure~\ref{fig:sccwalk4l}) are very good, bar the occasional troublesome graph.
It is clear that the maximum time given could be reduced from 1000s for many smaller graph sizes, with relatively few needing more than 100s.

\begin{figure}[H]
\centering
\includegraphics[scale=0.65]{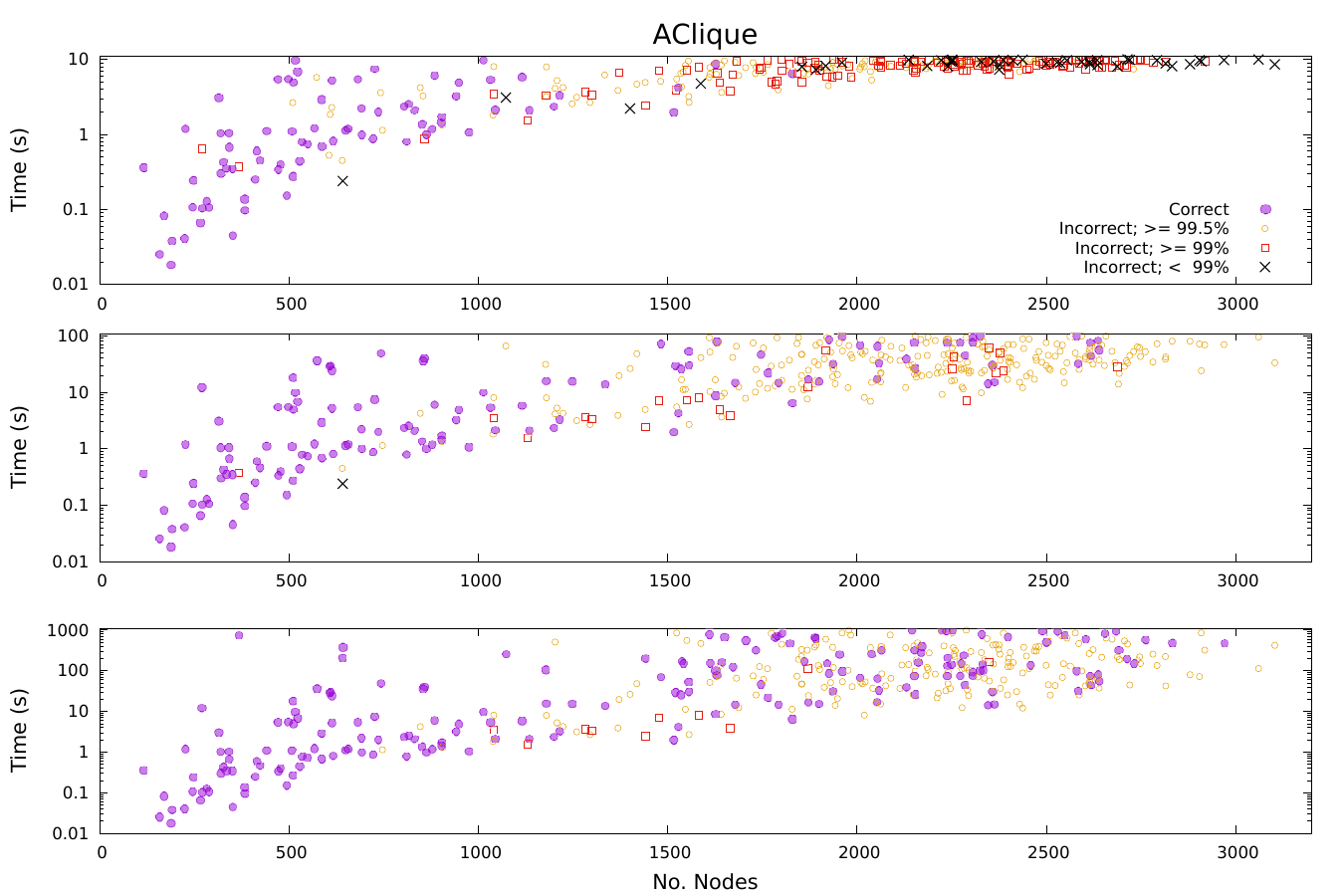}
\caption{Benchmarks of our own approximate solver \emph{AClique}. This returns multiple cliques after a given maximum time limit. We chart the results for 10, 100 and 1000 seconds of time. Colours represent the quality of result, showing perfect maximum cliques, imperfect cliques found of size $\geq 99.5\%$, cliques found of size between 99\% and 99.5\%, and cliques found below 99\%. Times shown are the time to find the best solution, which is typically considerably less than the specified execution time.}\label{fig:aclique}
\end{figure}

The results of our own genetic algorithm based clique finder, \emph{AClique}, are shown in Figure~\ref{fig:aclique}. This uses of a population of 96 cliques which are mutated and mated at random, along with randomised hill-climbing techniques, to build up the population. We periodically introduce new random cliques to avoid local maxima. This is purely random sampling and iterative improvements, with no branch and bound techniques. Despite this performance is in between \emph{FastWCLQ2} and \emph{SCCWalk4L} for time vs quality. While it struggles to find the correct maximum clique on many larger graphs it is rarely far away. However the key benefit is returning 100s or 1000s of cliques of the same maximum size found in a single execution run. We could combine many repeated runs of \emph{SCCWalk4L} with different random seeds to produce a similar result, perhaps even better, but we have not evaluated this and compare instead against \emph{AClique}.

\end{document}